\documentclass[fleqn,usenatbib]{mnras}

\usepackage{newtxtext,newtxmath}

\usepackage[T1]{fontenc}
\usepackage{ae,aecompl}

\usepackage{graphicx}	
\usepackage{amsmath}	
\usepackage{amssymb}	

\title[Explosion below the Stellar Surface]{Optical Transient from an Explosion Close to the Stellar Surface}

\author[Almog Yalinewich]{
Almog Yalinewich,$^{1}$\thanks{E-mail: almog.yalin@gmail.com}
Christopher D. Matzner$^{2}$
\\
$^{1}$Canadian Institute for Theoretical Astrophysics, 60 St. George St., Toronto, ON M5S 3H8, Canada\\
$^{2}$Department of Astronomy and Astrophysics, University of Toronto, 50 St. George Street, Toronto, ON M5S 3H4, Canada\\
}

\date{Accepted XXX. Received YYY; in original form ZZZ}

\pubyear{2015}

\begin{document}
\label{firstpage}
\pagerange{\pageref{firstpage}--\pageref{lastpage}}
\maketitle

\begin{abstract}
We study the hydrodynamic evolution of an explosion close to the stellar surface, and give predictions for the radiation from such an event. We show that such an event will give rise to a multi-wavelength transient. We apply this model to describe a precursor burst to the peculiar supernova iPTF14hls, which occurred in 1954, sixty year before the supernova. We propose that the new generation of optical surveys might detect similar transients, and they can be used to identify supernova progenitors well before the explosion.
\end{abstract}

\begin{keywords}
shock waves -- radiation: dynamics -- binaries: close
\end{keywords}



\section{Introduction}
One of the long standing mysteries about some types of supernovae is their progenitor \citep{Hirschi2017ProgenitorsSupernovae, Maeda2016ProgenitorsSupernovae}. The main problem in determining the progenitor is that by the time the supernova is brightest, the progenitor is already destroyed.

One way to obtain information about the progenitor is to consider radiative shock breakout. This is a brief and faint precursor to the main emission from the supernova. Shock breakout has been extensively studied in the past under the assumption of spherical symmetry \citep{Nakar2010EarlyBreakout, Sapir2013Non-relativisticBreakout}. These models predict that a small fraction of the stellar mass close to the stellar edge is accelerated to velocities two orders of magnitude larger than that of the bulk, and can emit hard photons due being outside blackbody thermodynamic equilibrium. However, it has been suggested that an oblique shock breakout would suppress these effects \citep{Matzner2013ObliqueImplications}. Unfortunately, to date only a handful of works studied the hydrodynamics of oblique breakout \citep[e.g.][]{Couch2008AsphericalJets, Couch2010Aspherical2008D} and even fewer accounted for non equilibrium radiation transfer \citep[e.g.][]{Afsariardchi2018AsphericalCurves}. We note that the source of the asymmetry in those previous studied are bipolar jets.

Different lines of evidence suggest that supernova explosions are asymmetrical. Measurements of the degree of polarisation indicate that core collapse supernovae are asymmetric \citep{Mazzali2005Astronomy:Bursts,Shapiro1982TheSymmetry,Leonard2006A2004dj}. Another indirect evidence for the asymmetry are neutron star natal kicks \citep{Beniamini2016FormationObservations, Tauris2017FormationSystems}. Similar observational evidence suggests that type Ia supernova are asymmetric \citep{Maeda2010AsymmetricSignatures}, and some theoretical models rely on non spherically symmetric processes to trigger the explosion. These models include white dwarf collisions \citep{Kushnir2013Head-onSupernovae}, mergers \citep{Raskin2014TypeDetonations} and instabilities \citep{Glasner2015IgnitionEnvelopes}.

Recently, it has been suggested that explosions can occur during common envelope phase. The explosion is triggered by a merger between the neutron star companion and the core of the progenitor, which causes it to release a large amount of energy in the form of jets \citep{Soker2017ExplainingSupernova}. The mechanism was invoked to explain the transient iPTF14hls \citep{Arcavi2017EnergeticStar}. About sixty years before the supernova, in 1954, a -16 magnitude burst was observed from the progenitor. It was suggested that this burst occurred upon entry of the neutron star companion into the progenitor envelope. In such a case, the explosion would have been close to the surface of the progenitor.

This scenario motivates us to consider an explosion that, instead of occurring at the centre of a star, occurs very close to the surface -- so close, in fact, that the radius of the star can be neglected.  Prior to breakout, the explosion evolves similarly a superbubble blowing out of a galactic disk \citep{Koo1990DynamicsMass, Koo1992DynamicsTheoryb}.
The explosion excavates a crater on the stellar surface and expels material from the basin. The explosion also generates thermal radiation, which is reprocessed by the ejecta before reaching an observer. A similar approach was used to describe the emission from the impact of comet Shoemaker Levy on Jupiter \citep{Zahnle1995AImpact}. 

\section{Domain of Validity}

In this work we concern ourselves with a certain transient that occurs when an amount of energy is released close to a stellar edge. Depending on the parameters of the event, qualitatively different outcomes can arise. In this section we discuss the requirements on the particular type of transient we have in mind.

First, the want the shock wave to not be relativistic when it approaches the stellar edge. This condition is given by
\begin{equation}
    \tilde{\Gamma} = \frac{E}{\rho_r l^3 c^2} < 1
\end{equation}
where $E$ is the energy of the explosion, $l$ is the distance of the hotspot from the stellar edge, $\rho_r$ is the density there and $c$ is the speed of light. A small fraction of the mass might still be accelerated to relativistic velocities \citep{Tan2000Trans-RelativisticProgenitors,Nakar2012RELATIVISTICSUPERNOVAE}, however, this will not affect the bulk material.

Second, we want the shock to be radiation dominated. This condition is equivalent to the requirement that the number of photons exceeds the number of particles, or that the matter dominated temperature is larger than the radiation dominated temperature. Matter dominated temperature is given by
\begin{equation}
    k T_m \approx m_p \frac{E}{\rho_r l^3}
\end{equation}
where $m_p$ is the proton mass. Radiation dominated temperature is given by
\begin{equation}
    k T_r \approx \left(E \frac{\hbar^3 c^3}{l^3}\right)^{1/4}
\end{equation}
where $\hbar$ is the reduced Planck constant. Hence, for radiation to dominate
\begin{equation}
    \tilde{\Gamma}^3 /\tilde{N} = \left(\frac{m_p}{\rho_r l^3}\right)^4 \left(\frac{E l}{c \hbar}\right)^3 > 1
\end{equation}
where $\tilde{N} = \frac{\rho_r \lambda_p^3}{m_p}$ is the number of baryons in a cube whose side is equal to the proton Compton wavelength $\lambda_p = \hbar/m_p c$.

Finally, we want the energy injection to be deep enough so that photons don't diffuse out instead of interacting with matter particles. This condition is equivalent to the requirement that the Sedov time of the explosion is shorter than the diffusion time. This condition is given by
\begin{equation}
    \tilde{\tau}^2 \tilde{\Gamma} = \frac{\kappa^2 \rho_r E}{l c^2} > 1
\end{equation}
where $\kappa$ is the opacity and $\tilde{\tau} = \kappa \rho_r l$.

When the shock reaches a distance comparable to $l$, we assume that it is moving much faster than the escape velocity from the host star, so we neglect gravity. However, as the shock expands and decelerates, then eventually the shock velocity will drop to the value of the escape velocity. When this happens, the shock truns into an acoustic wave and crater excavation stops.

Finally, we note that in this work we always assume that the material is fully ionised and that the dominant absorption process is Thompson scattering. Both assumptions break down when the temperature drops below $\sim 10^3 K$ and the gas begins to recombine.

\section{Early Breakout Evolution} \label{sec:early}

\subsection{Shock Ascent and Breakout}
In this section we consider a stellar explosion initiated by injecting an energy $E$ into a hotspot at a distance $l$ from the stellar surface. We further assume that the distance from the hotspot to the stellar surface is much smaller than the radius of the star $l \ll R_s$. The density profile is given by

\begin{equation}
    \rho_a = \rho_r \left(x/l\right)^{\omega}
\end{equation}
where $x$ is the depth below the surface, $\rho_r$ is the density there and $\omega$ is a constant. In a stellar atmosphere, the reference density would be $\rho_r \approx \frac{M_s}{R_s^3} \left(l/R_s\right)^{\omega}$, where $M_s$ is the mass of the star and $R_s$ is its radius. As a result of the energy injection, a shock wave will emerge from the hot spot and move outward. While the radius of the explosion is much smaller than the distance to the edge $l$ it will evolve as a Sedov Taylor explosion \citep{Sedov1946PropagationWaves, Taylor1950TheDiscussion}. When the explosion reaches a distance $l$ the mass swept by the shock is $\rho_r l^3$, and the shock velocity at that point is $\sqrt{E / \rho_r l^3}$. Afterwards, the upper part of the shock propagates into a declining density profile, and accelerates. The acceleration follows Sakurai's law $v \propto \rho^{-\mu}$, where $\mu \approx 0.19$ \citep{Sakurai1960OnGas}. Using the end of the Sedov Taylor phase as initial conditions, we obtain the ascending shock velocity

\begin{equation}
    v_a \approx \sqrt{\frac{E}{\rho_r l^3}} \left(\frac{x}{l}\right)^{-\mu \omega} \, .
\end{equation}

The acceleration stops when diffusion becomes faster than hydrodynamic advection \citep{Nakar2010EarlyBreakout}. This happens when the optical depth is comparable to the ratio between the speed of light and the material velocity
\begin{equation}
\tau \approx c/v_a \, . \label{eq:breakout_cond}
\end{equation}

Using criterion \ref{eq:breakout_cond} we can find the depth from which photons begin to break out
\begin{equation}
    \frac{x_{bo}}{l} \approx \left(\tilde{\tau} \sqrt{\tilde{\Gamma}}\right)^{-\frac{1}{1+\omega-\mu \omega}} \, .
\end{equation}
We call this depth the breakout shell. The velocity of the breakout shell is
\begin{equation}
    \frac{v_{bo}}{c} \approx \tilde{\Gamma}^{-\frac{\omega+1}{2 \left(1+\omega-\omega \mu\right)}} \tilde{\tau}^{\frac{\mu \omega}{1+\omega-\mu \omega}} \, .
\end{equation}
The density of the breakout shell is
\begin{equation}
    \frac{\rho_{bo}}{\rho_r} \approx \left(\tilde{\tau} \sqrt{\tilde{\Gamma}}\right)^{-\frac{\omega}{1+\omega - \mu \omega}} \, .
\end{equation}
The energy of the breakout shell is given by
\begin{equation}
    \frac{e_{bo}}{E} \approx \left(\tilde{\tau} \sqrt{\tilde{\Gamma}}\right)^{-1+\frac{\mu \omega}{1 + \omega - \mu \omega}} \, .
\end{equation}
The diffusion time at breakout is
\begin{equation}
    \frac{t_{bo}}{l/c} \approx \tilde{\Gamma}^{-\frac{1+\omega/2}{1+\omega - \mu \omega}} \tilde{\tau}^{-\frac{1+\mu \omega}{1+ \omega - \mu \omega}}
\end{equation}
whereas the light travel time across the breakout region is of order $l/c$.
Thus, assuming $t_{bo}>l/c$, at the very early stage of the shock breakout the luminosity rises to a value
\begin{equation}
    \frac{L_{bo}}{E c/l} \approx \tilde{\Gamma}^{\frac{1/2+\mu \omega}{1+\omega - \mu \omega}} \tilde{\tau}^{-\frac{1- 3 \mu}{1+ \omega - \mu \omega}}
\end{equation}

\subsection{Planar Phase}
In the next phase of the shock breakout, material expands only in a direction perpendicular to the stellar surface. This is called the planar phase. Sine the motion is slab-symmetric, the optical depth does not change, and so the luminosity is solely due to the adiabatically cooling breakout shell
\begin{equation}
    \frac{L_{pl}}{E c/l} \approx \frac{L_{bo}}{E c/l} \left(\frac{t_{bo}}{t}\right)^{4/3} \approx \tilde{\Gamma}^{\frac{6 \mu \omega - 4 \omega - 5}{6 \left(- \mu \omega + \omega + 1\right)}} \tilde{\tau}^{\frac{5 \mu \omega - 3 \omega - 4}{3 \left(- \mu \omega + \omega + 1\right)}} \tilde{t}^{-\frac{4}{3}}
\end{equation}
where $\tilde{t} = t \cdot c/l$.
This phase ends when the ejecta expands to a distance comparable to $l$, so the end of the planar phase occurs at time
\begin{equation}
    \frac{t_{pl}}{l/c} \approx \frac{l}{v_{bo}} / \frac{l}{c} \approx \tilde{\Gamma}^{-\frac{\omega +1}{2\left(1 + \omega - \mu \omega\right)}} \tilde{\tau}^{-\frac{\mu \omega}{1 + \omega - \mu \omega}}\, .
\end{equation}
The emitting area in this phase is $l^2$.

\subsection{Spherical Phase}
Next comes the spherical phase. In this phase shells expand spherically. The optical depth drops light from shells below the breakout shell emerges. We can calculate the density above the stellar surface by assuming the velocity of a fluid element does not change after it has been shocked and that the velocity profile is homologous, i.e.
\begin{equation}
    v_{h} \approx \frac{r}{t}
\end{equation}
where $r \gg l$ is the distance from the breakout site. A fluid element at position $r$ at time $t$ originated at depth
\begin{equation}
    \frac{x_{sp}}{l} \approx \left(\sqrt{\tilde{\Gamma}} \frac{\tilde{t}}{\tilde{r}}\right)^{\frac{1}{\mu \omega}}\, .
\end{equation}
The original mass that is now spread over a volume $r^3$ is
\begin{equation}
    \frac{m_{sp}}{\rho_r l^3} \approx \left(\sqrt{\tilde{\Gamma}}\frac{\tilde{t}}{\tilde{r}}\right)^{\frac{\omega+1}{\mu \omega}}
\end{equation}
and so the density is
\begin{equation}
    \frac{\rho_{sp}}{\rho_r} \approx \tilde{r}^{-3} \left(\sqrt{\Gamma}\frac{\tilde{t}}{\tilde{r}}\right)^{\frac{\omega+1}{\mu \omega}} \, .
\end{equation}
Photons diffuse from a radius where condition \ref{eq:breakout_cond} is satisfied. We call this position the luminosity shell, and its radius is given by
\begin{equation}
    \frac{r_{l,sp}}{l} \approx \tilde{\Gamma}^{-\frac{1+\omega}{2 \left(1 + \omega + \mu \omega\right)}} \tilde{\tau}^{\frac{\mu \omega}{1+\omega +\mu \omega}} \tilde{t}^{\frac{1 + \omega - \mu \omega}{1+ \omega + \mu \omega}}
\end{equation}
A fluid element currently at $r_{l,sp}$ originated from a depth $x_{l,sp} \approx x_{sp} \left(r=r_{l,sp}\right)$. The original energy of that shell was $\rho_a \left(x=x_{l,sp}\right) v_a \left(x=x_{l,sp}\right)^2 l^2 x_{l,sp}$ but by the time it has travelled a distance $r_{l,sp}$, adiabatic cooling reduced the energy by a factor of $\left(\frac{l^2 x_{l,sp}}{r_{l,sp}^3}\right)^{1/3}$. The diffusion timescale at the luminosity shell is comparable to the dynamical time $t$ and so the luminosity is given by
\begin{equation}
    \frac{L_{sp}}{E c /l} \approx \tilde{\Gamma}^{\frac{- \mu \omega + \frac{1}{6}}{\mu \omega + \omega + 1}} \tilde{\tau}^{\frac{\mu \omega - \omega - \frac{4}{3}}{\mu \omega + \omega + 1}} \tilde{t}^{\frac{- 4 \mu \omega + \frac{2}{3}}{\mu \omega + \omega + 1}}
\end{equation} 
The photospheric radius can be obtained by solving $\kappa \rho_{sp} r \approx 1$
\begin{equation}
    \frac{r_{p,sp}}{l} \approx \tilde{\tau}^{\frac{\mu \omega}{2 \mu \omega + \omega + 1}} \left(\sqrt{\tilde{\Gamma}} \tilde{t}\right)^{\frac{\omega + 1}{2 \mu \omega + \omega + 1}} \, .
\end{equation}
The spherical phase ends when the original depth of material at the luminosity shell is comparable with the initial depth of the hotspot $x_{l,sp} \approx l$
\begin{equation}
    \frac{t_{sp}}{l/c} \approx \frac{\sqrt{\tilde{\tau}}}{\sqrt[4]{\tilde{\Gamma}}}
\end{equation}

At later times the ejecta is dominated by material from the basin of the crater the explosion excavated. The radius of the crater grows as some powerlaw in time $R_c \propto t^{\beta}$, where the power law index is bounded by values corresponding to momentum and energy conservation \citep{Zeldovich1967PhysicsPhenomena}
\begin{equation}
    \frac{2}{5 + \omega} > \beta > \frac{1}{\omega+4} \, .
\end{equation}
In the next sections we perform numerical simulations to determine $\beta$ and calculate the lightcurve in the final phase.

\section{Crater Evolution} \label{sec:shock_trajectory}

\subsection{Adiabatic Evolution}

\begin{figure}
\includegraphics[width=0.9\columnwidth]{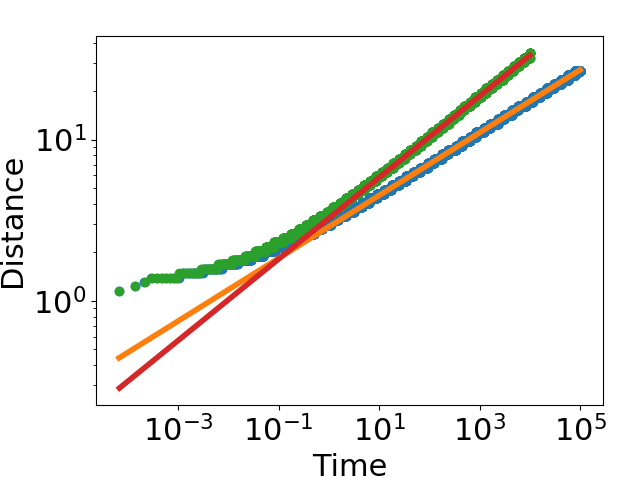}
\caption{
Shock trajectories obtained from simulations. The blue dots are for a radiative atmosphere ($\omega = 3$) and the green dots are for an adiabatic atmosphere ($\omega = 3/2$). The orange and red lines are power law fits to late times. The time and distance have been scaled according to the initial conditions: the distance scale is the depth of the explosion, and the velocity is scaled such that it is equal to 1 at the moment the top of the explosion reaches the stellar edge. The slope of the line in the case of a radiative atmosphere is $\beta \approx 0.19$, and in the case of an adiabatic atmosphere $\beta \approx 0.25$
}
\label{fig:shock_trajectory}
\end{figure}

\begin{figure}
\includegraphics[width=0.9\columnwidth]{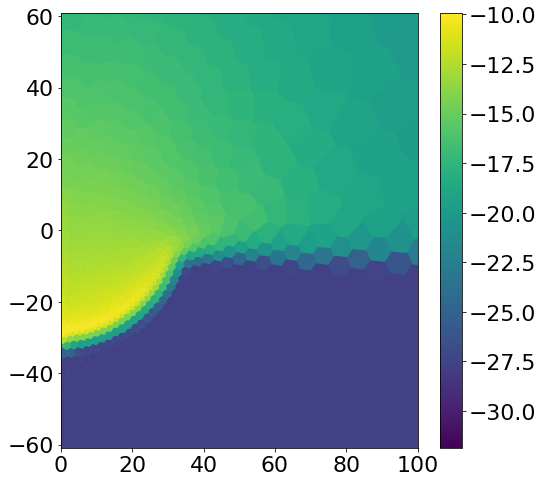}
\includegraphics[width=0.9\columnwidth]{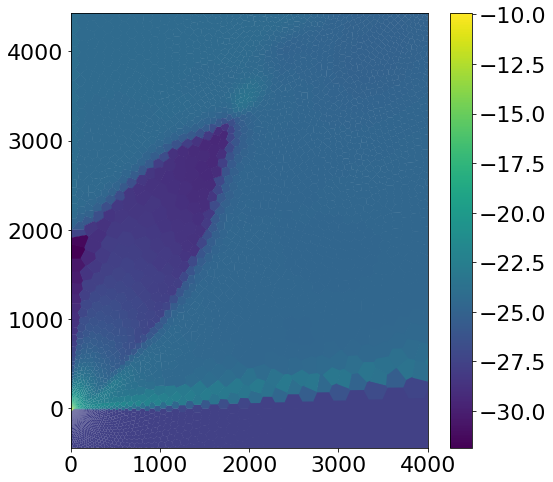}
\caption{
Log Pressure snapshot of an evolved crater. The top panel zooms in on the basin, while the bottom shows the outflow above the stellar surface. The top part ideally should not contain material, but since our simulation cannot handle vacuum, we had to fill this space with low density gas (mock vacuum). The dark wedge next to the origin represents outflow from the basin, while the bright semicircle is shocked mock vacuum. The polygons represent computational cells.
}
\label{fig:snapshot}
\end{figure}

To obtain the power law index for the crater radius with respect to time $d \ln R_c / d \ln t = \beta$, we ran a numerical simulation using the moving mesh hydrodynamic code RICH \citep{Yalinewich2015Rich:Mesh}. We considered two cases, one where the density scales with $\omega = 3$, corresponding to a radiative atmosphere, and $\omega = 3/2$, corresponding to an adiabatic atmosphere. In each case, we tracked the position of the shock wave at every time step, and fit the late time evolution to a power law. The results are shown in figure \ref{fig:shock_trajectory}. For a radiative atmosphere we obtain $\beta = 0.19$, and for the adiabatic atmosphere we obtain $\beta = 0.25$. A snapshot from the last time step of the one of the simulations is shown in figure \ref{fig:snapshot}.

\begin{figure}
\includegraphics[width=0.9\columnwidth]{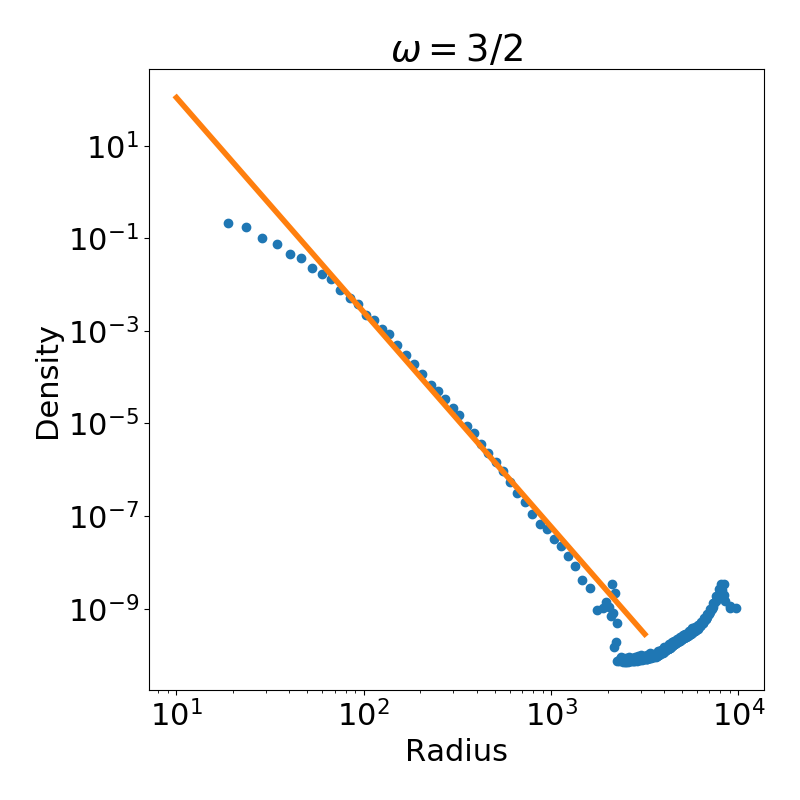}
\includegraphics[width=0.9\columnwidth]{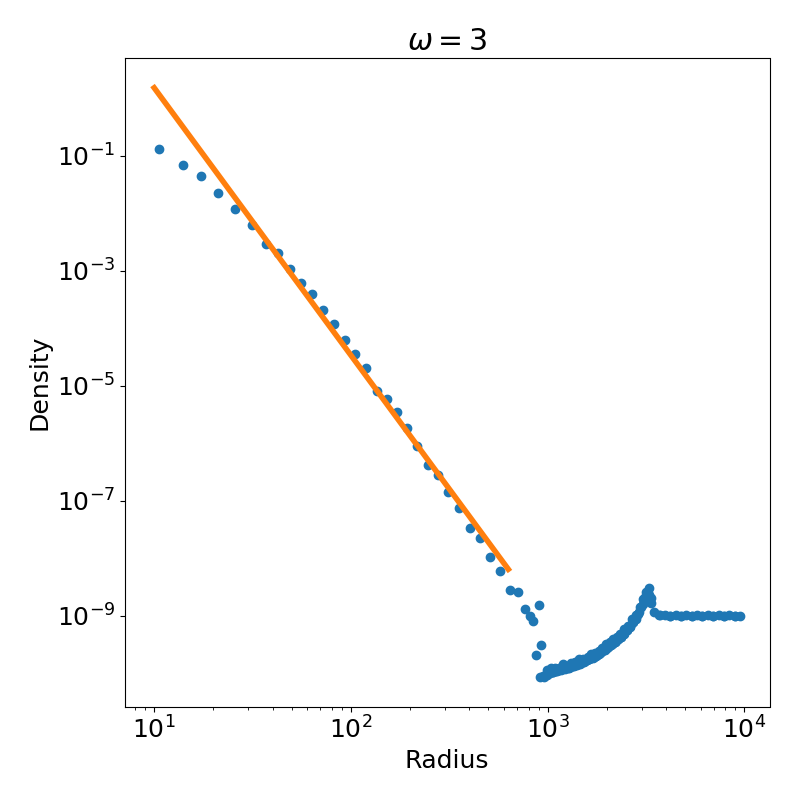}
\caption{
Profile of the density as a function of the distance from the hotspot, taken from adiabatic numerical simulations, for an adiabatic atmosphere (top) and a radiative atmosphere (bottom). The adiabatic index of the gas is $\gamma=5/3$ in both cases. The numerical results are in blue, and the power law fit is in orange. The theoretical power law indices (equation \ref{eq:crater_outer_density}) for the adiabatic atmosphere is -4.5 and for the radiative case it is -4.4. From the numerical simulation we can about -4.6 for both cases, so the numerical simulation roughly agrees with the theoretical model.}
\label{fig:density_profile}
\end{figure}

Now that we have $\beta$, we can obtain the crater radius as a function of time by connecting the crater power law to the early Sedov Taylor solution
\begin{equation}
    R_c \approx l \left(t \sqrt{\frac{E}{\rho_r l^5}}\right)^{\beta}
\end{equation}
From this relation we can also obtain a relation between the shock velocity and the crater size
\begin{equation}
    v_c \approx \sqrt{\frac{E}{\rho_r l^3}} \left(\frac{R_c}{l}\right)^{1-\frac{1}{\beta}}
\end{equation}
We can determine the density profile above the stellar surface in the phase where the basin is exposed in the same way as we did in the spherical phase. We assume a homologous velocity distribution in the ejecta, and that the velocity of a fluid element does not change after it is shocked. A fluid element in position $r$ at time $t$ originated from a depth of 
\begin{equation}
    \frac{x_{ac}}{l} \approx \left(\frac{\tilde{r}}{\sqrt{\tilde{\Gamma}} \tilde{t}} \right)^{\frac{\beta}{1-\beta}} \, .
\end{equation}
The density outside the stellar surface is
\begin{equation}
    \frac{\rho_{ac}}{\rho_r} \approx \tilde{r}^{-\frac{\beta \omega + 3}{1-\beta}} \left(\sqrt{\tilde{\Gamma}} \tilde{t}\right)^{\frac{\beta \left(\omega+3\right)}{1-\beta}} \, . \label{eq:crater_outer_density}
\end{equation}
To verify equation \ref{eq:crater_outer_density} we plotted the density as a function of radius from the simulations in figure \ref{fig:density_profile}. The power law index obtained from the simulation is very close to the theoretical value.
The luminosity radius is found using condition \ref{eq:breakout_cond}
\begin{equation}
    \frac{r_{l,ac}}{l} \approx \tilde{\Gamma}^{\frac{\beta \left(\omega + 3\right)}{2 \left(\beta \omega + 2 \beta + 1\right)}} \tilde{\tau}^{\frac{- \beta + 1}{\beta \omega + 2 \beta + 1}} \tilde{t}^{\frac{\beta \left(\omega + 4\right) - 1}{\beta \omega + 2 \beta + 1}} \, .
\end{equation}
The bolometric luminosity is given by
\begin{equation}
    \frac{L_{ac}}{E l/c} \approx \tilde{\tau}^{\frac{- 5 \beta - \omega \beta + 1}{2 \beta + \omega \beta  + 1}} \left(\sqrt{\tilde{\Gamma}} \tilde{t}^{2}\right)^{\frac{3 \beta - 2}{2 \beta + \omega \beta   + 1}}
\end{equation}
The photospheric can be obtained by solving $\kappa \rho_{ac} r \approx 1$, yielding
\begin{equation}
    \frac{r_{p,ac}}{l} \approx \tilde{\tau}^{\frac{- \beta + 1}{\beta \omega + \beta + 2}} \left(\sqrt{\tilde{\Gamma}} \tilde{t}\right)^{\frac{\beta \left(\omega + 3\right)}{\beta \omega + \beta + 2}} \, .
\end{equation}

We notice that the crater radius grows faster than the luminosity radius. For example, in the case of the adiabatic atmosphere, $d \ln R_c = d \ln t = 0.25$, whereas $d \ln r_{l,ac} / d \ln t = 0.2$, and in the case of a radiative atmosphere $d \ln R_c = d \ln t = 0.19$ whereas $d \ln r_{l,ac} / d \ln t = 0.17$. Both radii become comparable at time
\begin{equation}
    \frac{t_{ac}}{l/c} \approx \tilde{\Gamma}^{\frac{\beta \left(- \omega - 2\right)}{2 \left(\beta \omega + 2 \beta - 1\right)}} \tilde{\tau}^{- \frac{1}{\beta \omega + 2 \beta - 1}} \label{eq:adiabatic_crater_end_time}
\end{equation}
After time $t_{ac}$ radiation escapes directly from the crater basin. The evolution of the crater and the escaping radiation will be discussed in the next section.

\subsection{Radiative Evolution}

Since radiation leaks directly from the crater basin, it doesn't experience adiabatic losses as in the previous cases, so the luminosity is equal to the kinetic luminosity of the crater shock. However, because of the same reason the shock decelerates faster, since it only conserves momentum. This phase is analogous to the snowplough phase in supernova remnants. The shock wave is still self similar, but now with a power law index $\beta = \frac{1}{4 + \omega}$. The shock trajectory in this phase can be obtained by connecting it to the adiabatic phase.
\begin{equation}
    R_c \approx l \left(t_{ac} \sqrt{\frac{E}{\rho_r l^5}}\right)^{\beta} \left(\frac{t}{t_{ac}}\right)^{\frac{1}{\omega+4}}
\end{equation}
The kinetic luminosity is therefore
\begin{equation}
    \frac{L_{rc}}{El/c} \approx \tilde{\Gamma}^{\frac{- \beta \omega^{2} - 7 \beta \omega - 13 \beta + \omega + 4}{\beta \omega^{2} + 6 \beta \omega + 8 \beta - \omega - 4}} \tilde{\tau}^{\frac{- \beta \omega^{2} - 9 \beta \omega - 20 \beta + \omega + 5}{\beta \omega^{2} + 6 \beta \omega + 8 \beta - \omega - 4}} \tilde{t}^{- \frac{2 \omega + 7}{\omega + 4}} \, .
\end{equation}
We note that the system enters this last phase after a very long timescale (equation \ref{eq:adiabatic_crater_end_time}). In reality, crater excavation stops when the explosion pressure drops below the ambient pressure. This is probably happen before the system enters the radiative crater phase, so it may never menifest itself in a realistic scenario. The important results from the five stages discussed so far are summarised in table \ref{tab:phase_summary}.

\section{Radiation}

\subsection{Steady State Shock}

Let us consider a steady state shock wave moving at $v$ relative to some cold medium with density $\rho$. Let us further assume that the opacity of the medium is zero for the unshocked medium and constant for the shocked medium. In this configuration photons can escape from the shocked region and seen by an outside observer. We are interested in the average energy of those photons as a function of the shock velocity and density of the ambient medium. We will refer to this energy as the photon temperature, even though the temperature might not be strictly defined in cases where the system is not in thermal equilibrium.

At very low velocities the temperature is matter dominated, in which case
\begin{equation}
    k T_m \approx m_p v^2
\end{equation}
where $k$ is the Boltzmann constant. At higher velocities matter becomes radiation dominated, and then the temperature is given by
\begin{equation}
    k T_r \approx \left(\rho v^2 \hbar^3 c^3\right)^{1/4}
\end{equation}
where $\hbar$ is the reduced Planck constant. The transition from matter to radiation dominated shocks occurs when 
\begin{equation}
    \frac{v}{c} \approx \left(\frac{\rho \lambda_p^3}{m_p}\right)^{1/6} \approx 3 \cdot 10^{-3} \left(\frac{\rho}{1 \, \rm g/cm^3}\right)^{1/6}
\end{equation}
where $\lambda_p = \hbar / m_p c$ is the proton Compton wavelength. At even higher velocities the emerging photons are not in blackbody equilibrium with the shocked material. This is because only photons emitted close to the shock front can escape and make it to the observer. These photons are produced primarily via thermal Bremsstrahlung, and at high velocities this processes fails to produce enough photons to reach a blackbody thermal equilibrium. In this case, the temperature rises as a very steep function of the velocity, since as the velocity increases there is more thermal energy, and less photons to share it. In this photon starved regime the temperature is given by \citep{Katz2010FASTBREAKOUTS}
\begin{equation}
    k T_{ps} \approx m_e c^2 \left(\frac{m_p}{\alpha m_e}\right)^{2} \left(\frac{v}{c}\right)^8
\end{equation}
where $\alpha$ is the fine structure constant. The transition from the radiative to the photon starved regime happens when
\begin{equation}
    \frac{v}{c} \approx \left(\frac{\alpha m_e}{m_p}\right)^{4/15} \left(\frac{\rho \lambda_e^3}{m_e}\right)^{1/30} \approx 3 \cdot 10^{-2} \left(\frac{\rho}{1 \, \rm g/cm^3}\right)^{1/30}
\end{equation}
where $\lambda_e \approx \hbar / m_e c$ is the electron's Compton wavelength.  At even higher velocities pair production kicks in and prevents the temperature from exceeding pair production enevery
\begin{equation}
    k T_{pp} \approx m_e c^2 \, .
\end{equation}
The transition from the photon starved to the pair production regime happen when
\begin{equation}
    \frac{v}{c} \approx \left(\frac{\alpha m_e}{m_p}\right)^{1/4} \approx 4 \cdot 10^{-2} \, .
\end{equation}
At even higher velocities the temperature in the fluid frame remains at this constant value, but due to special relativistic effects the temperature observed increases as the velocity approaches the speed of light. In this paper we will not take into account relativistic effects.

\subsection{Evolving Shock}

In the previous section we saw that at when the shock is faster than about $10^4 \, \rm km/s$ then the emerging photons are not in blackbody thermal equilibrium. In this section we argue that non equilibrium emission is only important for the early phases (i.e. planar and spherical phases) and not for the crater phase. The main reason for that is that in the early phases the shock velocity changes very slowly with distance from the edge $v \propto x^{-\mu \omega}$, while in the cratering phase the velocity declines very steeply $v \propto x^{1-1/\beta}$, where $\beta \approx 0.2$. Therefore, even if the radiation is not in thermal equilibrium at the beginning of the crater phase, the shock velocity will quickly drop to a point where radiation is in blackbody equilibrium. In the late stages of the transient, in which is most likely to be observed, it will always be in blackbody thermal equilibrium.

\begin{table*}
\begin{tabular}{llll}
\hline
Phase            & Luminosity [$E c/l$] & Duration [$l/c$] & Photospheric radius [$l$] \\ \hline
Breakout & $\tilde{\Gamma}^{\frac{1/2+\mu \omega}{1+\omega - \mu \omega}} \tilde{\tau}^{-\frac{1- 3 \mu}{1+ \omega - \mu \omega}}$ & $\tilde{\Gamma}^{-\frac{1+\omega/2}{1+\omega - \mu \omega}} \tilde{\tau}^{-\frac{1+\mu \omega}{1+ \omega - \mu \omega}}$ & 1 \\
Planar            & $\tilde{\Gamma}^{\frac{6 \mu \omega - 4 \omega - 5}{6 \left(- \mu \omega + \omega + 1\right)}} \tilde{\tau}^{\frac{5 \mu \omega - 3 \omega - 4}{3 \left(- \mu \omega + \omega + 1\right)}} \tilde{t}^{-\frac{4}{3}}$           &     $\tilde{\Gamma}^{-\frac{\omega +1}{2\left(1 + \omega - \mu \omega\right)}} \tilde{\tau}^{-\frac{\mu \omega}{1 + \omega - \mu \omega}}$     &    1                 \\
Spherical        &      $\tilde{\Gamma}^{\frac{- \mu \omega + \frac{1}{6}}{\mu \omega + \omega + 1}} \tilde{\tau}^{\frac{\mu \omega - \omega - \frac{4}{3}}{\mu \omega + \omega + 1}} \tilde{t}^{\frac{- 4 \mu \omega + \frac{2}{3}}{\mu \omega + \omega + 1}}$      &   $\sqrt{\tilde{\tau}}/\sqrt[4]{\tilde{\Gamma}}$       &   $\tilde{\tau}^{\frac{\mu \omega}{2 \mu \omega + \omega + 1}} \left(\sqrt{\tilde{\Gamma}} \tilde{t}\right)^{\frac{\omega + 1}{2 \mu \omega + \omega + 1}}$                  \\
Adiabatic Crater &    $\tilde{\tau}^{\frac{- 5 \beta - \omega \beta + 1}{2 \beta + \omega \beta  + 1}} \left(\sqrt{\tilde{\Gamma}} \tilde{t}^{2}\right)^{\frac{3 \beta - 2}{2 \beta + \omega \beta   + 1}}$        &   $\tilde{\Gamma}^{\frac{\beta \left(- \omega - 2\right)}{2 \left(\beta \omega + 2 \beta - 1\right)}} \tilde{\tau}^{- \frac{1}{\beta \omega + 2 \beta - 1}}$       &        $\tilde{\tau}^{\frac{- \beta + 1}{\beta \omega + \beta + 2}} \left(\sqrt{\tilde{\Gamma}} \tilde{t}\right)^{\frac{\beta \left(\omega + 3\right)}{\beta \omega + \beta + 2}}$             \\
Radiative Crater &      $\tilde{\Gamma}^{\frac{- \beta \omega^{2} - 7 \beta \omega - 13 \beta + \omega + 4}{\beta \omega^{2} + 6 \beta \omega + 8 \beta - \omega - 4}} \tilde{\tau}^{\frac{- \beta \omega^{2} - 9 \beta \omega - 20 \beta + \omega + 5}{\beta \omega^{2} + 6 \beta \omega + 8 \beta - \omega - 4}} \tilde{t}^{- \frac{2 \omega + 7}{\omega + 4}}$      &    $\cdots$      &      $\left(\tilde{\tau} \tilde{t} \sqrt{\tilde{\Gamma}}\right)^{1/3}$               \\ \hline
\end{tabular}
\caption{The luminosity, duration and photospheric radius for each of the four phases discussed in sections \ref{sec:early} and \ref{sec:shock_trajectory}} \label{tab:phase_summary}
\end{table*}

\section{Application} \label{sec:application}

\begin{figure}
    \includegraphics[width=\linewidth]{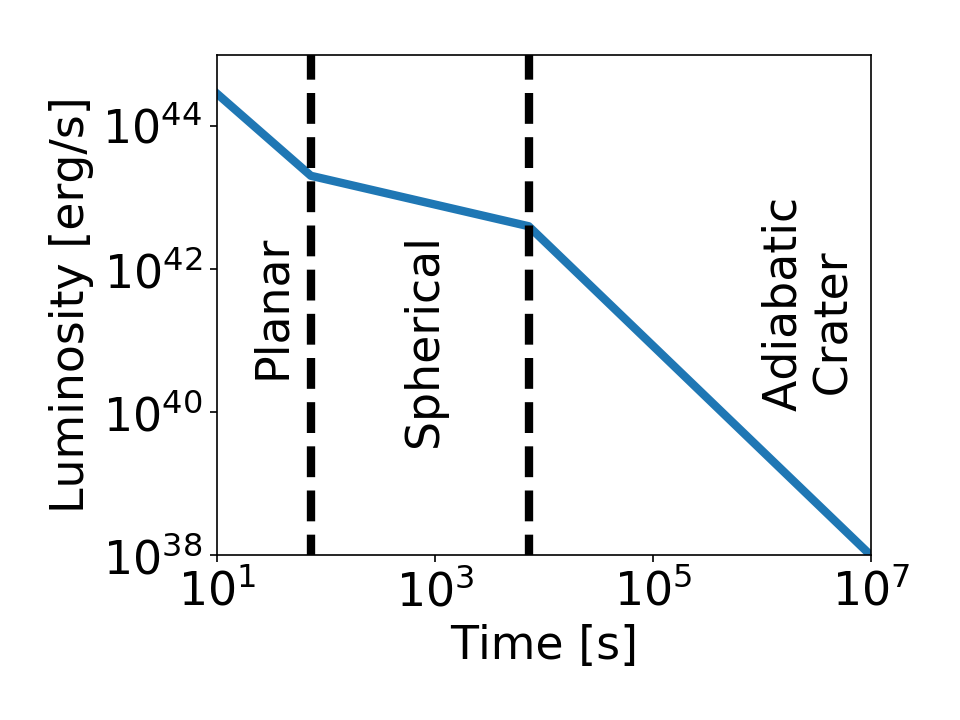}
\caption{Bolometric light curve for the explosion described in section \ref{sec:application}. The hydrodynamic phase is indicated next to each segment.}
\label{fig:bolometric_lightcurve}
\end{figure}

\begin{figure}
    \includegraphics[width=\linewidth]{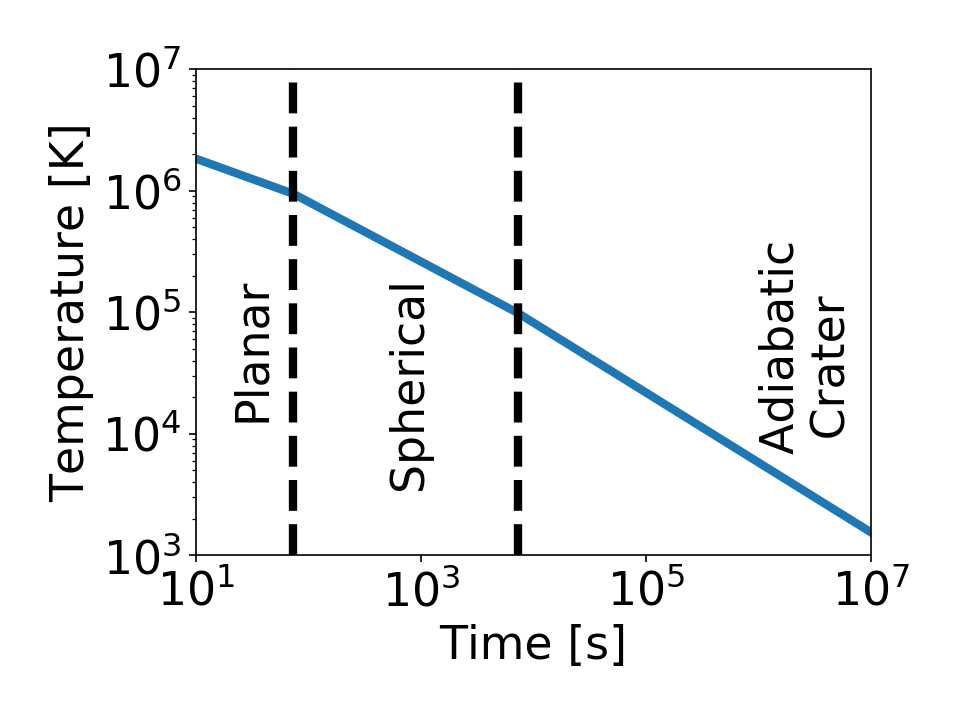}
\caption{Temperature evolution in the case described in section \ref{sec:application}. The hydrodynamic phase is indicated next to each segment.}
\label{fig:temperature}
\end{figure}

In this section we consider the burst observed in 1954 from the progenitor of iPTF14hls. According to the model of \citep{Soker2017ExplainingSupernova}, the 1954 burst occurred when the companion neutron star first entered the envelope of the progenitor star - a $80 M_{\odot}$ supergiant with a radius of $R_s \approx 100 R_{\odot}$. Due to the techonological capabilities at the time, only the absolute magnitude of the event at peak is known: about -16 in the r band \citep{Arcavi2017EnergeticStar}, which roughly translates to a luminosity of $10^8 L_{\odot} \approx 10^{41} \, \rm erg/s$. Assuming this is the luminosity at the end of the spherical phase (or the beginning of the exposed basin phase), a radiative envelope $\omega=3$, and and that the explosion occurred in a moderate depth $l/R_s \approx 0.1$, we can estimate the energy to be of the order of $E \approx 10^{48} \, \rm erg$. Using these estimates, and the discussion from the previous sections, we plotted the theoretical lightcurve in figure \ref{fig:bolometric_lightcurve} and the evolution of the blackbody temperature in figure \ref{fig:temperature}. We see that in the early stages of the explosion (i.e. the planar and spherical phases) the temperature was in the mild X-ray and UV range, so it would not be observable from earth. At the beginning of the crater phase the temperature drops to the visible range. We note that in the planar and spherical phases the radiation can depart from blackbody equilibrium, in which case the temperature might be even higher, and so the prospect of observing these phases from earth is even lower.

It is reasonable to assume that if such a burst would be detected today, then the duration and temperature would also be measured. Using the extra data it would be possible to also infer the depth at which the explosion occurred.

\section{Conclusions}

In this work calculated the lightcurves and temperature from an explosion close to the surface of a star. The early evolution of such an explosion follows the same phases as in the case of a planar breakout. However, this transient has a third, long lasting phase where radiation leaks from a shock that excavates a crater in the stellar atmosphere.

We use this model to describe the 1954 outburst of the progenitor iPTF14hls \citep{Arcavi2017EnergeticStar}. According to the theoretical model of \citep{Soker2017ExplainingSupernova}, iPTF14hls is the result of a merger between a companion neutron star and the core of a giant star, and the 1954 burst occurs during the initial plunge of the neutron star into the giant's envelop. We calibrate the energy of the burst according to the observed luminosity, and from it we calculate lightcurves and a temperature. We find that in the early phases the temperature was too large to be observed from earth, but in the crater phase the temperature drops to the visible range.

With the launch of next generation optical surveys, like ZTF \citep{Bellm2017TheSky} and LSST \citep{Robertson2017LargeRoadmap}, it would be easier to detect similar transients in the future. Proper identification and modelling of these transients, could, in principle, be considered as a precursor to the supernova explosion. This would allow us to study the properties of the progenitor antemortem.

\section*{Acknowledgements}

AY would like to thank Re'em Sari, Ehud Nakar, Noam Soker and James Guillochon for  useful discussions. AY is supported by the Vincent and Beatrice Tremaine Fellowship. CDM's research is supported by an NSERC discovery grant. This work made use of the sympy \citep{Meurer2017SymPy:Python} and matplotlib \citep{Hunter2007Matplotlib:Environment} python packages.




\bibliographystyle{mnras}
\bibliography{references} 

\begin{thebibliography}{}
\makeatletter
\relax
\def\mn@urlcharsother{\let\do\@makeother \do\$\do\&\do\#\do\^\do\_\do\%\do\~}
\def\mn@doi{\begingroup\mn@urlcharsother \@ifnextchar [ {\mn@doi@}
  {\mn@doi@[]}}
\def\mn@doi@[#1]#2{\def\@tempa{#1}\ifx\@tempa\@empty \href
  {http://dx.doi.org/#2} {doi:#2}\else \href {http://dx.doi.org/#2} {#1}\fi
  \endgroup}
\def\mn@eprint#1#2{\mn@eprint@#1:#2::\@nil}
\def\mn@eprint@arXiv#1{\href {http://arxiv.org/abs/#1} {{\tt arXiv:#1}}}
\def\mn@eprint@dblp#1{\href {http://dblp.uni-trier.de/rec/bibtex/#1.xml}
  {dblp:#1}}
\def\mn@eprint@#1:#2:#3:#4\@nil{\def\@tempa {#1}\def\@tempb {#2}\def\@tempc
  {#3}\ifx \@tempc \@empty \let \@tempc \@tempb \let \@tempb \@tempa \fi \ifx
  \@tempb \@empty \def\@tempb {arXiv}\fi \@ifundefined
  {mn@eprint@\@tempb}{\@tempb:\@tempc}{\expandafter \expandafter \csname
  mn@eprint@\@tempb\endcsname \expandafter{\@tempc}}}

\bibitem[\protect\citeauthoryear{Afsariardchi \& Matzner}{Afsariardchi \&
  Matzner}{2018}]{Afsariardchi2018AsphericalCurves}
Afsariardchi N.,  Matzner C.~D.,  2018, \mn@doi [The Astrophysical Journal,
  Volume 856, Issue 2, article id. 146, 19 pp. (2018).]
  {10.3847/1538-4357/aab3d4}, 856

\bibitem[\protect\citeauthoryear{Arcavi et~al.,}{Arcavi
  et~al.}{2017}]{Arcavi2017EnergeticStar}
Arcavi I.,  et~al., 2017, \mn@doi [Nature, Volume 551, Issue 7679, pp. 210-213
  (2017).] {10.1038/nature24030}, 551, 210

\bibitem[\protect\citeauthoryear{Bellm \& Kulkarni}{Bellm \&
  Kulkarni}{2017}]{Bellm2017TheSky}
Bellm E.~C.,  Kulkarni S.~R.,  2017, \mn@doi [Nature Astronomy, Volume 1, id.
  0071 (2017).] {10.1038/s41550-017-0071}, 1

\bibitem[\protect\citeauthoryear{Beniamini \& Piran}{Beniamini \&
  Piran}{2016}]{Beniamini2016FormationObservations}
Beniamini P.,  Piran T.,  2016, \mn@doi [Monthly Notices of the Royal
  Astronomical Society] {10.1093/mnras/stv2903}

\bibitem[\protect\citeauthoryear{Couch, Wheeler  \& Milosavljevic}{Couch
  et~al.}{2008}]{Couch2008AsphericalJets}
Couch S.~M.,  Wheeler J.~C.,   Milosavljevic M.,  2008, \mn@doi [The
  Astrophysical Journal, Volume 696, Issue 1, pp. 953-970 (2009).]
  {10.1088/0004-637X/696/1/953}, 696, 953

\bibitem[\protect\citeauthoryear{Couch, Pooley, Wheeler  \&
  Milosavljevic}{Couch et~al.}{2010}]{Couch2010Aspherical2008D}
Couch S.~M.,  Pooley D.,  Wheeler J.~C.,   Milosavljevic M.,  2010, \mn@doi
  [The Astrophysical Journal, Volume 727, Issue 2, article id. 104, 16 pp.
  (2011).] {10.1088/0004-637X/727/2/104}, 727

\bibitem[\protect\citeauthoryear{Glasner, Livne, Steinberg, Yalinewich  \&
  Truran}{Glasner et~al.}{2015}]{Glasner2015IgnitionEnvelopes}
Glasner S.~A.,  Livne E.,  Steinberg E.,  Yalinewich A.,   Truran J.~W.,  2015,
  MNRAS

\bibitem[\protect\citeauthoryear{Hirschi, Arnett, Cristini, Georgy, Meakin  \&
  Walkington}{Hirschi et~al.}{2017}]{Hirschi2017ProgenitorsSupernovae}
Hirschi R.,  Arnett D.,  Cristini A.,  Georgy C.,  Meakin C.,   Walkington I.,
  2017, \mn@doi [Proceedings of the International Astronomical Union]
  {10.1017/S1743921317004896}

\bibitem[\protect\citeauthoryear{Hunter}{Hunter}{2007}]{Hunter2007Matplotlib:Environment}
Hunter J.~D.,  2007, \mn@doi [Computing in Science and Engineering]
  {10.1109/MCSE.2007.55}, 9, 90

\bibitem[\protect\citeauthoryear{Katz, Budnik  \& Waxman}{Katz
  et~al.}{2010}]{Katz2010FASTBREAKOUTS}
Katz B.,  Budnik R.,   Waxman E.,  2010, \mn@doi [The Astrophysical Journal]
  {10.1088/0004-637X/716/1/781}, 716, 781

\bibitem[\protect\citeauthoryear{Koo \& McKee}{Koo \&
  McKee}{1990}]{Koo1990DynamicsMass}
Koo B.-C.,  McKee C.~F.,  1990, \mn@doi [The Astrophysical Journal]
  {10.1086/168712}, 354, 513

\bibitem[\protect\citeauthoryear{Koo \& McKee}{Koo \&
  McKee}{1992}]{Koo1992DynamicsTheoryb}
Koo B.-C.,  McKee C.~F.,  1992, \mn@doi [The Astrophysical Journal]
  {10.1086/171133}, 388, 103

\bibitem[\protect\citeauthoryear{Kushnir, Katz, Dong, Livne  \&
  Fern{\'{a}}ndez}{Kushnir et~al.}{2013}]{Kushnir2013Head-onSupernovae}
Kushnir D.,  Katz B.,  Dong S.,  Livne E.,   Fern{\'{a}}ndez R.,  2013, \mn@doi
  [Astrophysical Journal Letters] {10.1088/2041-8205/778/2/L37}

\bibitem[\protect\citeauthoryear{Leonard et~al.,}{Leonard
  et~al.}{2006}]{Leonard2006A2004dj}
Leonard D.~C.,  et~al., 2006, \mn@doi [Nature] {10.1038/nature04558}

\bibitem[\protect\citeauthoryear{Maeda}{Maeda}{2010}]{Maeda2010AsymmetricSignatures}
Maeda K.,  2010, in Susa H.,  Arnould M.,  Gales S.,  Motobayashi T.,
  Scheidenberger C.,   Utsunomiya H.,  eds,  American Institute of Physics
  Conference Series Vol. 1238, American Institute of Physics Conference Series.
  pp 163--168, \mn@doi{10.1063/1.3455921}

\bibitem[\protect\citeauthoryear{Maeda \& Terada}{Maeda \&
  Terada}{2016}]{Maeda2016ProgenitorsSupernovae}
Maeda K.,  Terada Y.,  2016, \mn@doi [International Journal of Modern Physics
  D, Volume 25, Issue 10, id. 1630024] {10.1142/S021827181630024X}

\bibitem[\protect\citeauthoryear{Matzner, Levin  \& Ro}{Matzner
  et~al.}{2013}]{Matzner2013ObliqueImplications}
Matzner C.~D.,  Levin Y.,   Ro S.,  2013, \mn@doi [Astrophysical Journal]
  {10.1088/0004-637X/779/1/60}

\bibitem[\protect\citeauthoryear{Mazzali et~al.,}{Mazzali
  et~al.}{2005}]{Mazzali2005Astronomy:Bursts}
Mazzali P.~A.,  et~al., 2005, \mn@doi [Science] {10.1126/science.1111384}

\bibitem[\protect\citeauthoryear{Meurer et~al.,}{Meurer
  et~al.}{2017}]{Meurer2017SymPy:Python}
Meurer A.,  et~al., 2017, \mn@doi [PeerJ Computer Science]
  {10.7717/peerj-cs.103}, 3, e103

\bibitem[\protect\citeauthoryear{Nakar \& Sari}{Nakar \&
  Sari}{2010}]{Nakar2010EarlyBreakout}
Nakar E.,  Sari R.,  2010, \mn@doi [Astrophysical Journal]
  {10.1088/0004-637X/725/1/904}

\bibitem[\protect\citeauthoryear{Nakar \& Sari}{Nakar \&
  Sari}{2012}]{Nakar2012RELATIVISTICSUPERNOVAE}
Nakar E.,  Sari R.,  2012, \mn@doi [The Astrophysical Journal]
  {10.1088/0004-637X/747/2/88}, 747, 88

\bibitem[\protect\citeauthoryear{Raskin, Kasen, Moll, Schwab  \&
  Woosley}{Raskin et~al.}{2014}]{Raskin2014TypeDetonations}
Raskin C.,  Kasen D.,  Moll R.,  Schwab J.,   Woosley S.,  2014, \mn@doi
  [Astrophysical Journal] {10.1088/0004-637X/788/1/75}

\bibitem[\protect\citeauthoryear{Robertson et~al.,}{Robertson
  et~al.}{2017}]{Robertson2017LargeRoadmap}
Robertson B.~E.,  et~al., 2017, eprint arXiv:1708.01617

\bibitem[\protect\citeauthoryear{Sakurai}{Sakurai}{1960}]{Sakurai1960OnGas}
Sakurai A.,  1960, Communications on Pure and Applied Mathematics, 13, 353

\bibitem[\protect\citeauthoryear{Sapir, Katz  \& Waxman}{Sapir
  et~al.}{2013}]{Sapir2013Non-relativisticBreakout}
Sapir N.,  Katz B.,   Waxman E.,  2013, \mn@doi [Astrophysical Journal]
  {10.1088/0004-637X/774/1/79}

\bibitem[\protect\citeauthoryear{Sedov}{Sedov}{1946}]{Sedov1946PropagationWaves}
Sedov L.~I.,  1946, Prikl. Mat. Mekh, 10, 241

\bibitem[\protect\citeauthoryear{Shapiro \& Sutherland}{Shapiro \&
  Sutherland}{1982}]{Shapiro1982TheSymmetry}
Shapiro P.~R.,  Sutherland P.~G.,  1982, \mn@doi [The Astrophysical Journal]
  {10.1086/160559}

\bibitem[\protect\citeauthoryear{Soker \& Gilkis}{Soker \&
  Gilkis}{2017}]{Soker2017ExplainingSupernova}
Soker N.,  Gilkis A.,  2017, \mn@doi [Monthly Notices of the Royal Astronomical
  Society, Volume 475, Issue 1, p.1198-1202] {10.1093/mnras/stx3287}, 475, 1198

\bibitem[\protect\citeauthoryear{Tan, Matzner  \& McKee}{Tan
  et~al.}{2000}]{Tan2000Trans-RelativisticProgenitors}
Tan J.~C.,  Matzner C.~D.,   McKee C.~F.,  2000, \mn@doi [The Astrophysical
  Journal] {10.1086/320245}, 551, 946

\bibitem[\protect\citeauthoryear{Tauris et~al.,}{Tauris
  et~al.}{2017}]{Tauris2017FormationSystems}
Tauris T.~M.,  et~al., 2017, ] {10.3847/1538-4357/aa7e89}, 846, 170

\bibitem[\protect\citeauthoryear{Taylor}{Taylor}{1950}]{Taylor1950TheDiscussion}
Taylor G.~I.,  1950, Proc. R. Soc. Lond. A, 201, 159

\bibitem[\protect\citeauthoryear{Yalinewich, Steinberg  \& Sari}{Yalinewich
  et~al.}{2015}]{Yalinewich2015Rich:Mesh}
Yalinewich A.,  Steinberg E.,   Sari R.,  2015, \mn@doi [Astrophysical Journal,
  Supplement Series] {10.1088/0067-0049/216/2/35}, 216

\bibitem[\protect\citeauthoryear{Zahnle \& Mac~Low}{Zahnle \&
  Mac~Low}{1995}]{Zahnle1995AImpact}
Zahnle K.,  Mac~Low M.-M.,  1995, \mn@doi [Journal of Geophysical Research]
  {10.1029/95JE01620}, 100, 16885

\bibitem[\protect\citeauthoryear{Zel'dovich \& Raizer}{Zel'dovich \&
  Raizer}{1967}]{Zeldovich1967PhysicsPhenomena}
Zel'dovich Y.~B.,  Raizer Y.~P.,  1967, {Physics of shock waves and
  high-temperature hydrodynamic phenomena}.
Dover

\makeatother
\end{thebibliography}








\bsp	
\label{lastpage}
\end{document}